\documentclass[12pt,lettersize]{article}

\begin{document}
	

\thispagestyle{empty}
	
\begin{center}
	{\bf \LARGE Phenomenologically viable gravitational theory based on a preferred foliation without extra modes}
	\vspace*{15mm}
		
	{\large Jorge Bellor\'{\i}n}$^{1}$
	\vspace{3ex}
		
	{\it Department of Physics, Universidad de Antofagasta, 1240000 Antofagasta, Chile.}
	\vspace{3ex}
		
	$^1${\tt jbellori@gmail.com} \hspace{1em}
		
	\vspace*{1em}
		
	\vspace*{15mm}
	{\bf Abstract}
	\begin{quotation}{\small\noindent
		We present a gravitational field theory that implements Ho\v{r}ava's proposal of foliation-preserving-diffeomorphisms symmetry and higher spatial curvature directly in the canonical formalism. Due to the higher spatial derivative the theory is potentially renormalizable. Since this gauge symmetry is natural in the canonical formalism, we do not require a Lagrangian of second-order in time derivatives to begin with. We define the nonzero part of the Hamiltonian and the constraints motivated by the kinetic-conformal version of the nonprojectable Ho\v{r}ava theory. The resulting theory is an extension of the latter, in the sense that it admits more solutions. Among the additional solutions there are homogeneous and isotropic configurations governed by the Friedmann equations. The theory has the same number of propagating degrees of freedom of general relativity. At the linearized level it reproduces the tensorial gravitational waves of general relativity. We discuss how observational bounds can be satisfied.
			}
	\end{quotation}
\end{center}

\thispagestyle{empty}

\paragraph{Introduction.}
Ho\v{r}ava theory \cite{Horava:2008ih,Horava:2009uw} is based on the FDiff (foliation-preserving diffeomorphisms) symmetry. In this scheme the gravitational space is composed by a foliation of spacelike hypersurfaces, and the foliation can not be changed since it has absolute physical meaning. The motivation \cite{Horava:2009uw} for doing this is to introduce higher-order spatial curvature terms that may render the theory renormalizable, avoiding the ghosts that tipically affect the relativistic higher-curvature theories \cite{Stelle:1976gc}.

In its generic formulation, due to the reduced gauge symmetry group, the Ho\v{r}ava theory propagates one physical mode in addition to the ones of General Relativity. However, there is an exception to this rule that does not introduce further symmetries. It happens when the coupling constant of the kinetic term, in the nonprojectable formulation, takes its critical value $1/d$, where $d$ is the spatial dimension \cite{Bellorin:2013zbp}. In this  critical case the kinetic term of the Lagrangian acquires an anisotropic conformal symmetry \cite{Horava:2009uw}. We call this case the kinetic-conformal Ho\v{r}ava theory, and we label it as the kcH-theory for short. In the kcH-theory there are two extra second-class constraints that suprime the extra mode. The second-class constraints are not directly related to gauge symmetries (but still there are some relations between the kcH-theory and the anisotropc conformal Ho\v{r}ava theory \cite{Bellorin:2018blt}, since the former can be seen as the explicit breaking of the conformal symmetry of the later). The matching in degrees of freedom between the kcH-theory and General Relativity is not limitated to a mere coincidence in number. At the linearized level, the dynamics of effective theory for large distances of the kcH-theory coincides with the dynamics of linearized General Relativity: it propagates gravitational waves of the same tensorial modes \cite{Bellorin:2013zbp}. This is interesting since there is strong evidence favouring that the observed gravitational waves corresponds to tensorial polarizations, at least for the case of pure polarizations \cite{Abbott:2017oio,Abbott:2018lct}. Another version of the Ho\v{r}ava theory that does not propagate the extra mode is the so-called $U(1)$ extension proposed in Ref.~\cite{Horava:2010zj}. This is based on the projectable theory, and the extra mode is eliminated by means of a $U(1)$ gauge symmetry. 

The kcH-theory exhibits serious limitations when reproducing homogeneous and isotropic configurations. This is manifested as a divergence in the effective gravitational constant of cosmological scale \cite{Blas:2009qj,Bellorin:2017gzj}. Despite of this we think that the rest of the phenomenology that has been studied for the kcH-theory is interesting, see \cite{Bellorin:2016hcu}. We highlight again that the theories based on the FDiff symmetry and higher spatial curvature are potentially renormalizable and unitary. Indeed, the renormalizability of the projectable Ho\v{r}ava theory has been shown in Ref.~\cite{Barvinsky:2015kil}.

Our interest is to look for a modification of the kcH-theory that might enlarge the space of solutions, admiting in particular more cosmological configurations. We want to preserve the fundamental characteristic of the kcH-theory: the same number of physical degrees of freedom of general relativity. To find such a theory (or model), we first realize that the canonical action is completely FDiff-covariant by itself. Thus, we may apply Ho\v{r}ava's ideas directly in the canonical formalism, without requiring a Lagrangian of second order in time derivatives as the starting point. Instead, we define the nonzero part of the Hamiltonian and the constraints, defining in this way the action of the theory. These ingredients are motivated by the ones of the kcH-theory, but with modifications that allow for more physically interesting configurations.

Following Refs.~\cite{Horava:2009uw,Blas:2009qj}, the theory has a potential of sixth order in spatial derivatives, including all the inequivalent terms that are compatible with the FDiff symmetry. Although we do not deal with the full potential explicitly, we scrutinize the effective theory for large distances in three ways: perturbative equations for the propagation of gravitational waves, homogeneous and isotropic configurations, and the observational bounds coming from gravitational waves and the PPN (parameterized-post-Newtonian) parameters. The perturbative analysis is an important test of consistency, since it makes transparent the set of field equations that are explicitly solved as elliptic equations and the ones that remain hiperbolic.

In the homogeneous and isotropic case we find a set of Friedmann equations. This is promisory since one may expect predictions at cosmological scale near to general relativity. We give in advance that the observational bounds fix strongly the values of two of the coupling constants of the large-distance effective theory to their corresponding relativistic values (this also happens in the kcH-theory \cite{Bellorin:2016hcu}). This requeriment is not in contradiction with the fundamental features of the theory. We also comment that the higher-derivative terms are not considered in the tests of the observational bounds.


\paragraph{FDiff symmetry in canonical formalism.}
\label{sec:invariantaction}
We study the general form of the FDiff-invariant canonical action whose canonical variables are the pair $(g_{ij},\pi^{ij})$ and the lapse function $N$, which has vanishing canonically conjugated momentum. This last condition is a constraint of the theory and, since it is an already-solved constraint, we reduce the phase space by putting this canonical momentum equal to zero everywhere in the action. The general form the canonical action is
\begin{equation}
S = 
\int dt d^3x \left[ \pi^{ij} \dot{g}_{ij} 
- \left( \mathcal{H}_0 + N_i \mathcal{H}^i + B_A \phi^A \right) \right] \,,
\label{generalcanonicalaction}
\end{equation}
where $\mathcal{H}_0$ is the ``nonzero" part of the Hamiltonian density, that is, the part that is not proportional to constraints, hence it remains nonzero in the totally reduced phase space. $\mathcal{H}^i$ is the momentum constraint,
\begin{equation}
\mathcal{H}^i \equiv - 2 \nabla_k \pi^{ik} \,,
\label{momentumconstraint}
\end{equation}
and $\phi^A$ stands for the rest of constraints. $N_i$ and $B_A$ enter as Lagrange multipliers. All the standard notation of Riemannian manifolds is referred to the spatial metric $g_{ij}$: spatial indices are raised and lowered with it, $\nabla_i$ uses its Levi-Civita connection, and so on.

Given a system of coordinates $(\vec{x},t)$ on the foliation, the coordinate transformations that preserve the foliation are defined by $\delta t = f(t)$, $\delta x^i = \zeta^i(\vec{x},t)$. As gauge transformations of the field variables, the corresponding FDiff transformations are given by (changing signs of $f$ and $\zeta^i$)
\begin{eqnarray}
&& \delta g_{ij} = 
\zeta^k \partial_k g_{ij} + \partial_i \zeta^k g_{kj}  
+ \partial_j \zeta^k g_{ik} + f \dot{g}_{ij}  \,,
\label{gtransformation}
\\ &&
\delta \pi^{ij} = 
\zeta^k \partial_k \pi^{ij} - \partial_k \zeta^i \pi^{kj} 
- \partial_k \zeta^j \pi^{ik}  + \partial_k \zeta^k \pi^{ij}  
+ f \dot{\pi}^{ij} \,,
\label{pitransformation}
\\ &&
\delta N = \zeta^k \partial_k N + f \dot{N} + \dot{f} N \,,
\label{ntransformation}
\\ &&
\delta N_i = 
\zeta^k \partial_k N_i + \partial_i \zeta^k N_k 
+ \dot{\zeta}^k g_{ik} + f \dot{N}_i + \dot{f} N_i \,,
\label{nitransformation}
\\ &&
\delta B_A = 
\zeta^k \partial_k B_A + f\dot{B}_A + \dot{f} B_A \,.
\label{sigmatransformation}
\end{eqnarray}
The gauge transformations of $g_{ij}$, $N$ and $N_i$ are taken from the Ho\v{r}ava theory \cite{Horava:2009uw}, since they are the Arnowitt-Deser-Misner (ADM) variables of the foliation. In particular, the third term in the right-hand side of (\ref{nitransformation}) is relevant for the gauge transformation of the momentum constraint, $N_i$ being its Lagrange multiplier. The transformation of $\pi^{ij}$ (\ref{pitransformation}) characterizes it as a tensorial density under spatial diffeomorphisms and as a scalar under the transformation of time. Its transformation is designed to balance the kinetic term. Finally, we assume that the rest of Lagrange multipliers, $B_A$, transform as scalars under spatial diffeomorphisms and as densities under time transformations. For general FDiff, only the parameters $\zeta^i$ are arbitrary functions over the space and the time, whereas $f$ does not. Hence, in the strict sense, only the spatial diffeomorphisms, which are those given by $f = 0$ and general $\zeta^i$, are gauge transformations over the foliation.

The central aim of this section is to point out that the FDiff symmetry can be implemented directly on the canonical action, and that this hold for general $\mathcal{H}_0$ and $\phi^A$. To this end it is required that these objects transform as
\begin{eqnarray}
&&  \delta \mathcal{H}_0 = 
\zeta^k \partial_k \mathcal{H}_0 + \partial_k \zeta^k \mathcal{H}_0
+ f \dot{\mathcal{H}}_0 + \dot{f} \mathcal{H}_0  \,,
\label{deltah0}
\\ &&
\delta \phi^A = 
\zeta^k \partial_k \phi_A + \partial_k \zeta^k \phi_A + f\dot{\phi_A} \,.
\label{deltaphi}
\end{eqnarray}
That is, the $\phi^A$ must be scalar densities under spatial diffeomorphisms and scalars under time transformations, whereas $\mathcal{H}_0$ behaves as a density under both transformations (of course, one can be more general by asking the combination $B_A \phi^A$ to transform as a double density). These requeriments are easy to meet with the appropiated combinations of the canonical variables and their derivatives, as happens in the Hamiltonian formulation of the nonprojectable Ho\v{r}ava theory \cite{Kluson:2010nf,Donnelly:2011df,Bellorin:2011ff,Bellorin:2013zbp}. Under these conditions, the proof of the FDiff symmetry on the canonical action (\ref{generalcanonicalaction}) is rather straightforward since most of the objects involved in the action (\ref{generalcanonicalaction}) are tensors or tensor densities under FDiff. The less direct step is the balancing between the transformation of the kinetic term $\pi^{ij} \dot{g}_{ij}$ and the one of $N_i \mathcal{H}^i$ under a time-dependent spatial diffeomorphism, since from the transformation of the kinetic term there remains time derivatives of the gauge parameter,
\begin{equation}
\delta ( \pi^{ij} \dot{g}_{ij} ) \sim \pi^{ij} \partial_t \delta g_{ij} \sim 
\pi^{ij} \left( \dot{\zeta}^k \partial_k g_{ij} 
+ 2 \partial_i \dot{\zeta}^k g_{kj} \right)\,.
\end{equation}
These terms are compensated by the ones remainning from the transformation of $N_i \mathcal{H}^i$,
\begin{equation}
\delta ( N_i \mathcal{H}^i ) \sim \delta N_i \mathcal{H}^i \sim 
-2 \dot{\zeta}^k g_{ik} \nabla_j \pi^{ij} \,.
\end{equation}
This mechanisms happens exactly in the same way in the ADM formulation of general relativity, since the time-dependent spatial diffeomorphisms are part of its gauge symmetry. This is the reason why we separate the momentum constraint $\mathcal{H}^i$ from the rest of constraints and write explicitly it together with its Lagrange multiplier $N_i$. 


\paragraph{The gravitational theory.}
Adapting Ho\v{r}ava's ideas \cite{Horava:2009uw} to the canonical formalism, we start by setting the most general nonzero part of the Hamiltonian density, $\mathcal{H}_0$, that transforms under FDiff as (\ref{deltah0}), has separate kinetic and potential parts and is quadratic in the canonical momentum. It is
\begin{equation}
\mathcal{H}_0 = 
\frac{N}{\sqrt{g}} \left( \pi^{ij} \pi_{ij} - \omega \pi^2 \right) 
+ \sqrt{g} N \mathcal{V} \,,
\label{h0}
\end{equation}
where $\omega$ is an arbitrary dimensionless constant, $\pi \equiv g_{ij} \pi^{ij}$, and the potential $\mathcal{V}[g_{ij},a_i]$ is the most general scalar under general FDiff that can be written in terms of the spatial metric $g_{ij}$, the spatial vector $a_i = \partial_i \ln N$ \cite{Blas:2009qj} and their spatial derivatives. According to the discussion in \cite{Horava:2009uw}, $z=3$, where $2z$ is the highest order in spatial derivatives of $\mathcal{V}$, is the minimal order required for the power-counting renormalizability in $3$ spatial dimensions. We adopt this criterium, hence we consider that the potential $\mathcal{V}$ is of order $z=3$. In Ho\v{r}ava theory it is known that the number of terms in a $z=3$ potential of the nonprojectable theory is of order $10^2$, see for example Ref.~\cite{Blas:2009qj}. In this paper we deal with the potential $\mathcal{V}$ formally. We will present explicitly only the $z=1$ truncation, which gives the effective theory for large distances. We shall explore several physical features of this effective theory, as well as its theoretical consistency.

The next step is the definition of the constraints. We remark again that in the approach we are following the canonical theory is defined from the beginning, rather than obtained from a previous action. Hence, there is a big freedom to define the constraints explicitly. We use as guiadance the closeness to the kinetic-conformal Ho\v{r}ava theory, which in turn is close to general relativity (at least its large-distance effective theory). Thus, the first constraint we define, besides the momentum constraint $\mathcal{H}^i$ already included, is the condition
\begin{equation}
 \mathcal{H} \equiv \frac{\delta}{\delta N} \int d^3x \mathcal{H}_0 = 0 \,,
 \label{hamiltonianconstfunctional}
\end{equation}
which is similar to the way the so-called Hamiltonian constraint arises in general relativity and the Ho\v{r}ava theory. For the $\mathcal{H}_0$ given in (\ref{h0}), we obtain
\begin{eqnarray}
&&
\mathcal{H} = 
\frac{1}{\sqrt{g}} \left( \pi^{ij} \pi_{ij} - \omega \pi^2 \right) 
+ \mathcal{U} = 0 \,,
\label{hamiltonianconstraint}
\\ &&
\mathcal{U} \equiv 
\frac{\delta}{\delta N} \int d^3y \sqrt{g} N \mathcal{V} \,.
\label{uderivative}
\end{eqnarray}
Note that once the constraint (\ref{hamiltonianconstfunctional}) is imposed, the equation of motion derived from the variation of $N$ only gets contributions from the constraint sector of the canonical action, that is, from $B_A \phi^A$, since constraint (\ref{hamiltonianconstfunctional}) demands that the contribution coming from $\mathcal{H}_0$ must be zero. As a consequence, the evolution equation associated to $\delta_N$ is necessary homogeneous in the Lagrange multipliers $B_A$.

To define our second constraint, we observe that one of the constraints that characterizes the kcH-theory is the condition $\pi = 0$. We propose to replace it by
\begin{equation}
 \phi^1 \equiv \nabla^2 \pi = 0 \,.
 \label{nablapiconstraint}
\end{equation}
This form admits a bigger set of solutions; for example, on flat spatial geometries, any nonzero $\pi$ that depends only of time is a nontrivial solution of (\ref{nablapiconstraint}).

Our third and last constraint is also inspired by the kcH-theory. In that theory, starting with a second-order Lagrangian, the constraint $\pi = 0$ arises as a primary constraint. Consequently, one must impose its time preservation. Mimicking this, the last constraint we propose is inspired by the time preservation of (\ref{nablapiconstraint}). We define the constraint
\begin{equation}
 \phi^2 \equiv 
 \frac{1}{N} \left\{ \phi^1 , \int d^3x \mathcal{H}_0 \right\} = 0 \,, 
\end{equation}
where the prefactor $N^{-1}$ is introduced in order to meet (\ref{deltaphi}). For the $\mathcal{H}_0$ given in (\ref{h0}) we get
\begin{equation}
\phi^2 =
 N^{-1} \nabla^2 \mathcal{E}
= 0 \,,
\label{cconstraintpre}
\end{equation}
where
\begin{eqnarray}
&&
\mathcal{E} \equiv 
 N \left( \frac{3}{2} \mathcal{U} + \mathcal{W} \right) \,,
\label{e}
\\ &&
\mathcal{W}^{ij} \equiv 
\frac{1}{N} \frac{\delta}{\delta g_{ij}} \int d^3y \sqrt{g} N \mathcal{V}
\,, \quad
\mathcal{W} = g_{ij} \mathcal{W}^{ij} \,.
\label{wderivative} 
\end{eqnarray}
In the above we have used the constraint $\mathcal{H} = 0$  (\ref{hamiltonianconstraint}) explicitly to eliminate the $\pi^{ij}$-dependence of $\phi^2$. In summary, given the potential $\mathcal{V}$, the canonical action is given by
\begin{eqnarray}
&&
S = 
\int dt d^3x \left[ \pi^{ij} \dot{g}_{ij} - \left(
\mathcal{H}_0 + N_k \mathcal{H}^k 
+ B_1 \phi^1 + B_2 \phi^2 + B_3 \mathcal{H} \right)\right] \,,
\label{canonicalaction}
\\ &&
\mathcal{H}_0 = 
\frac{N}{\sqrt{g}} \left( \pi^{ij} \pi_{ij} - \omega \pi^2 \right) 
+ \sqrt{g} N \mathcal{V} \,,
\label{canonicalactionfin}
\end{eqnarray}
and the constraints are
\begin{eqnarray}
&&
\mathcal{H}^i = - 2 \nabla_k \pi^{ik} = 0\,,
\label{momentumconstraintfin}
\\ &&
\mathcal{H} = 
\frac{1}{\sqrt{g}} \left( \pi^{ij} \pi_{ij} - \omega \pi^2 \right) 
+ \mathcal{U} = 0 \,,
\\ &&
\phi^1 = \nabla^2 \pi = 0 \,,
\label{nablapi}
\\ && 
\phi^2 =
N^{-1} \nabla^2 \mathcal{E} = 0\,,
\label{cconstraint}
\end{eqnarray}
where $\mathcal{E}$ is given in (\ref{e}) and $\mathcal{U}$ in (\ref{uderivative}). According to the definitions of $\mathcal{H}_0$, the constraints and the transformation of the Lagrange multipliers, the theory possesses the FDiff gauge symmetry defined by (\ref{gtransformation}) - (\ref{sigmatransformation}).

The dynamical field variables of the theory are $g_{ij}$, $\pi^{ij}$ and $N$. The four constraints  $\mathcal{H}^i$, $\mathcal{H}$, $\phi^1$ and $\phi^2$ eliminate six functional degrees of freedom and there are three gauge degrees of freedom corresponding to the symmetry of the spatial diffeomorphisms. This leaves four independent physical degrees of freedom in the phase space, which corresponds to two even modes of propagation. There is not extra modes.

We make some remarks on the approach we have followed for defining the theory and the way the several variables must be found. In the literature of field theory, the canonical formulation of a given theory is usually obtained following the Dirac procedure, which starts with a Lagrangian as input with its associated primary constraints. Then, to extract further constraints in the canonical version the preservation on time of the primary constraints is required, and so on. Our approach here is different, since the canonical action is covariant by itself with respect to the symmetry of interest. We have started with the canonical action, defining the nonzero part of the Hamiltonian and the constraints in a way compatible with the FDiff symmetry.  This step is equivalent to the usual way of defining a theory by means of a Lagrangian of second (or higher) order in time derivatives.

Let us call the canonical variables and the Lagrange multipliers collectively as the field variables. Excluding the gauge degrees of freedom, the field variables must be solved from the full system of field equations obtained by taking variations of the action with respect to all of them. This point of view is based on the principle of stationary point of the action. In particular the Lagrange multipliers must be obtained in this way, excluding, again, their possible usage as gauge degrees of freedom. The Lagrange multipliers arise only in the field equations yielded by the variations with respect to $g_{ij}$ and $\pi^{ij}$, which yield evolution equations (involve $\dot{g}_{ij}$ and $\dot{\pi}^{ij}$), and the variation with respect to $N$, which does not yield explicit time derivative. We show these three equations for the effective theory in (\ref{gpuntoz1free}), (\ref{pipuntoz1free}) and (\ref{eqmultipliers}). The final physical and consistent content of the theory arises if solutions of the full system of field equations can be found. In the subsequent sections we show that this is the case in several scenarios. For each notrivial solution of the full set of field equations that is found, all the constraints are automatically preserved in time by the given configuration since the field equations, which include the constraints, are solved for all time and for each point of the spatial slices. Solutions obtained in this way need no more tests of consistency since they are stationary points of the action.

As a consequence of the reduction given by the elimination of the canonically conjugate momentum of the lapse function $N$, the action does not depend explicitly on $\dot{N}$. Despite this, $N$ in general acquires an induced dependence on time since it is one of the dynamical variables used to solve the field equations. Related to this is the fact that the Poisson bracket does not involve $N$. For instance, if a quantity $\Psi$ depends functionally on $N$, its time derivative receives a contribution proportional to $\dot{N}$ that must be added to the Poisson bracket defined in terms of derivatives with respect to the pair $(g_{ij},\pi^{ij})$. Explicitly,
\begin{equation}
 \dot{\Psi} = \left\{ \Psi, H \right\} + \frac{\delta \Psi}{\delta N} \dot{N} \,,
\end{equation}
where $H$ is the Hamiltonian. The last term can be eliminated if the reduction also includes $N$, that is, if $N$ is solved, for instance, from a constraint.

\paragraph{The kinetic-conformal Ho\v{r}ava theory is contained.}
The canonical action of the kinetic-conformal Ho\v{r}ava theory \cite{Bellorin:2013zbp}, once the momentum conjugated of $N$ is set to zero, is
\begin{eqnarray}
&&
\tilde{S} = 
\int dt d^3x \left[ \pi^{ij} \dot{g}_{ij} - \left(
\tilde{\mathcal{H}}_0 + N_k \mathcal{H}^k 
+ B_1 \tilde{\phi}^1 + B_2 \tilde{\phi}^2 + B_3 \tilde{\mathcal{H}} \right)\right] \,,
\\ &&
\tilde{\mathcal{H}}_0 = 
\frac{N}{\sqrt{g}} \pi^{ij} \pi_{ij} 
+ \sqrt{g} N \mathcal{V} \,,
\end{eqnarray}
and its constraints are
\begin{eqnarray}
&&
\tilde{\mathcal{H}} = 
\frac{1}{\sqrt{g}} \pi^{ij} \pi_{ij}  
+ \mathcal{U} = 0 \,,
\\ &&
\tilde{\phi}^1 = \pi = 0 \,,
\label{nablapi}
\\ && 
\tilde{\phi}^2 = \frac{3}{2} \mathcal{U} + \mathcal{W} = 0\,,
\label{cconstraint}
\end{eqnarray}
where $\mathcal{U}$ and $\mathcal{W}$ are defined in (\ref{uderivative}) and (\ref{wderivative}). Thus, the two differences between the kcH-theory and the theory we have in this paper, provided that the same potential $\mathcal{V}$ is used in both cases, are that the latter has the $\omega \pi^2$ term in $\mathcal{H}_0$ and $\mathcal{H}$ and that it has the Laplacian operator acting on the  $\phi^{1,2}$ constraints. The momentum constraint $\mathcal{H}^i$ is shared by both theories exactly in the same form due to the FDiff symmetry.

It turns out that any solution of the full set of constraints of the kcH-theory is a solution of the full set of the constraints of the theory we present here since conditions $\tilde{\phi}^{1,2} = 0$ imply $\phi^{1,2} = 0$, and $\tilde{\mathcal{H}}$ becomes equal to $\mathcal{H}$ due to the $\tilde{\phi}^1 = \pi = 0$ condition. Moreover, the canonical evolution equations of the kcH-theory, combined with its constraints, imply the canonical evolution equations of the  theory we present here. This is easy to see in terms of variations of the action. First, any contribution of the $\omega \pi^2$ term to the evolution equations vanishes if the constraint $\tilde{\phi}^1 = \pi = 0$ of the kcH-theory is satisfied. Second, the variations of the $\phi^{1,2}$ constraints have the schematic form (varying the metric, for example)
\begin{eqnarray}
&&
\delta ( B_1 \phi^1 ) 
\sim
B_1 (\delta \nabla^2) \pi + B_1 \nabla^2 \delta \pi \,,
\label{deltaphi1}
\\ &&
\delta ( B_2 \phi^2 ) 
\sim
B_2 N^{-1} (\delta \nabla^2) \mathcal{E} 
+ B_2 N^{-1} \nabla^2 \delta \mathcal{E}  \,,
\label{deltaphi2}
\end{eqnarray}
with the quantity $\mathcal{E}$ defined in (\ref{e}). If the constraints of the kcH-theory are satisfied, the terms $B_1 (\delta \nabla^2) \pi$ and $B_2 N^{-1} (\delta \nabla^2) \mathcal{E}$ are zero due to $\tilde{\phi}^{1,2}=0$. The terms $B_1 \nabla^2 \delta \pi$ and $B_2 N^{-1} \nabla^2 \delta \mathcal{E}$, after integration by parts, are equivalent to varying the $\tilde{\phi}^1$ and $\tilde{\phi}^2$ constraints of the kcH-theory accompained by the Lagrange multipliers $\nabla^2 B_1$ and $ \nabla^2 (B_2/N)$ respectively. Thus, if the constraints of the kcH-theory are satisfied, then the equations of motion of the kcH-theory are equal to the ones of this theory, after the appropiate redefinition of the Lagrange multipliers. Summarizing, we have that the constraints and the canonical equations of motion of this theory are implied by the constraints and the canonical equations of motion of the kcH-theory. Therefore, any solution of the kcH-theory is a solution of this theory -all the dynamics of the kcH-theory is contained in the theory we present here. The new thing is that the converse is not true in general, this theory admits more solutions.

\paragraph{Asymptotically flat configurations.}
There is another correspondence between this and the kcH-theory that goes in the opposite direction: a class of solutions of this theory that can be proven to be solutions of the kcH-theory. This is the important case of the asymptotically flat configurations. 

We use the standard definition of asymptotic flatness in canonical formalism \cite{Regge:1974zd},
\begin{equation}
g_{ij} = \delta_{ij} + \mathcal{O}(r^{-1}) \,,
\quad
\pi^{ij} = \mathcal{O}(r^{-2}) \,,
\quad
N = 1 + \mathcal{O}(r^{-1})
\label{asympflat}
\end{equation}
Let us denote by $M_{\mbox{\tiny AF}}$ the subspace of the phase space where the conditions (\ref{asympflat}) hold. It turns out that in $M_{\mbox{\tiny AF}}$ the only solutions to the conditions $\nabla^2 \pi = 0$ and $\nabla^2 \mathcal{E} = 0$ are $\pi = 0$ and $\mathcal{E} = 0$, yielding the constraints of the kcH-theory $\tilde{\phi}^{1,2}$, if we assume again that the same potential $\mathcal{V}$ is taken to define both theories. Let us show how this works in detail. At a given instant of time $t$, we multiply constraint $\phi^1$ by $\pi/\sqrt{g}$, integrate over the whole spatial hypersurface and then integrate by parts,
\begin{equation}
0 = 
\int \frac{d^3x}{\sqrt{g}} \pi \nabla^2 \pi
= 
\oint\limits_{\infty} d\Sigma_k \pi \partial^k \left( \frac{\pi}{\sqrt{g}} \right)  
- \int d^3x \sqrt{g} \partial_k \left( \frac{\pi}{\sqrt{g}} \right) 
\partial^k \left( \frac{\pi}{\sqrt{g}} \right) \,.
\label{piceropre}
\end{equation}
According to (\ref{asympflat}), the integrand of the surface integral at infinity is of order $\mathcal{O}(r^{-5})$, hence this integral vanishes. The integrand of the last integral is manifestly nonnegative. Assuming continuity of the integrand, the entire integral is zero if and only if the integrand is zero point to point. Since this is a condition on the Riemannian modulus of a three vector, the whole vector vanishes,
\begin{equation}
\partial_i \left( \frac{\pi}{\sqrt{g}} \right) = 0 \,.
\end{equation}
The general solution of this equation is that $\pi/\sqrt{g}$ is an arbitrary function of time, but the asymptotic conditions (\ref{asympflat}) demand that this function is equal to zero. Thus, in $M_{\mbox{\tiny AF}}$ we get that the constraint $\phi^1 = \nabla^2 \pi = 0$ is reduced to $\pi = 0$ for all $\vec{x},t$, which is the $\tilde{\phi}^1$ constraint of the kcH-theory. Constraint $\phi^2$ (\ref{cconstraintpre}) can be handled in a similar way: if we multiply it by $N \mathcal{E}/\sqrt{g}$ and integrate over all the spatial hypersurface, then we get its equivalent reduced form in $M_{\mbox{\tiny AF}}$: $\mathcal{E} = 0$ for all $\vec{x},t$, which is the $\tilde{\phi}^2$ constraint of the kcH-theory. Constraint $\mathcal{H}$ also coincides with $\tilde{\mathcal{H}}$ since the $\omega \pi^2$ term disappears in $M_{\mbox{\tiny AF}}$ (constant $\omega$ plays no role in $M_{\mbox{\tiny AF}}$). With regards to the equations of motion, the arguments are the same of the previous section, since once we have arrived at the conditions $\pi = \mathcal{E} = 0$ in $M_{\mbox{\tiny AF}}$, the same variational arguments (with the same redefinition of the Lagrange multipliers) show that the equations of motion of the kcH-theory are implied by the ones of this theory in $M_{\mbox{\tiny AF}}$. Therefore, all the asymptotically flat solutions of this theory are also solutions of the kcH-theory. Since in the previous section we have seen that any solution of the later is a solution of the former, we conclude that the set of asymptotically flat configurations is exactly the same for both theories.

\paragraph{Large-distance effective theory.}
To define the effective theory for large distances we truncate the potential $\mathcal{V}$ at the lowest order in spatial derivatives, leaving only the $z=1$ terms (here we do not consider cosmological constant). The $z=1$ potential is
\begin{equation}
 \mathcal{V}^{(z=1)} = - \beta R - \alpha a_k a^k \,,
\end{equation} 
where $\beta,\alpha$ are coupling constants. Thus, the nonzero part of the Hamiltonian takes the form
\begin{equation}
 \mathcal{H}_0 = 
 \frac{N}{\sqrt{g}} \left( \pi^{ij} \pi_{ij} - \omega \pi^2 \right) 
 - \sqrt{g} N \left( \beta R + \alpha a_k a^k \right) \,,
\end{equation}
and the set of constraints becomes
\begin{eqnarray}
&&
\mathcal{H}^i = - 2 \nabla_k \pi^{ik} = 0\,,
\label{momentumconst}
\\ &&
\mathcal{H} = 
\frac{1}{\sqrt{g}} \left( \pi^{ij} \pi_{ij} - \omega \pi^2 \right) 
- \sqrt{g} \left( \beta R + \alpha a_k a^k  - 2\alpha \frac{\nabla^2 N}{N}  \right)
	= 0 \,,
\\ &&
\phi^1 = \nabla^2 \pi = 0 \,,
\label{phiuno}
\\ && 
\phi^2 =
N^{-1} \nabla^2 \mathcal{E} = 0\,,
\end{eqnarray}
where
\begin{equation}
 \mathcal{E} = 
 \sqrt{g} N \left( \beta R + \alpha a_k a^k \right) 
 - \gamma_2 \sqrt{g} {\nabla^2 N} \,,
 \label{phi2z1} 
 \quad
 \gamma_2 \equiv \beta + {3\alpha}/{2}
\end{equation}
Constraints $\mathcal{H}^i$ and $\phi^1$ preserve their forms given in Eqs.~(\ref{momentumconstraint}) and (\ref{nablapiconstraint}) since they do not depend on the chosen potential. In the above we have multipled $\phi^2$ by $-1/2$.

The equations of motion obtained by taking variations of the action (\ref{canonicalaction}) with respect to $\pi^{ij}$, $g_{ij}$ and $N$ are, respectively,
\begin{eqnarray}
&&
\dot{g}_{ij} = 
\frac{2}{\sqrt{g}} \left( N + B_3 \right) 
\left( \pi_{ij} - \omega g_{ij} \pi \right) 
+ 2 \nabla_{(i} N_{j)} + g_{ij} \nabla^2 B_1 \,,
\label{gpuntoz1free}
\\ &&
\dot{\pi}^{ij} = 
- \frac{2}{\sqrt{g}} ( N + B_3 ) 
\left[ \pi^{ik} \pi_k{}^j - \omega \pi \pi^{ij} 
- \frac{1}{4} g^{ij} \left( \pi^{kl} \pi_{kl} - \omega \pi^2 \right) \right]
\nonumber
\\ &&
- 2 \nabla_k N^{(i} \pi^{j)k} + \nabla_k ( N^k \pi^{ij})
- \nabla^{(i} B_1 \nabla^{j)} \pi
+ \frac{1}{2} g^{ij} \nabla_k B_1 \nabla^k \pi
- \nabla^2 B_1 \left( \pi^{ij} - \frac{1}{2} g^{ij} \pi \right)
\nonumber
\\ &&
- \sqrt{g} \left[
\beta ( R^{ij} - \frac{1}{2} g^{ij} R )
+ \alpha ( a^i a^j - \frac{1}{2} g^{ij} a_k a^k ) 
- \beta ( \nabla^{ij} - g^{ij} \nabla^2 ) \right] ( N + B_3 )
\nonumber
\\ &&
+ \sqrt{g} \left(
\nabla^{(i} N \nabla^{j)}- \frac{1}{2} g^{ij} \nabla_k N \nabla^k  \right) \left( \gamma_2 \nabla^2 \tilde{B}_2 - 2 \alpha \tilde{B}_3 \right)
\nonumber
\\ &&
- \sqrt{g} \left[ 
\beta \left( \nabla^{ij} - g^{ij} \nabla^2 - R^{ij} \right)
- \alpha \sqrt{g} a^i a^j \right] (N \nabla^2 \tilde{B}_2)
\nonumber 
\\ &&
- \frac{\gamma_2}{2} \sqrt{g} g^{ij} \nabla^2 N \nabla^2 \tilde{B}_2
- \nabla^{(i} \tilde{B}_3 \nabla^{j)} \mathcal{E} 
+ \frac{1}{2} g^{ij} \nabla_k \tilde{B}_3 \nabla^k \mathcal{E} 
\,,
\label{pipuntoz1free}
\\ &&
0 = \gamma_2 \nabla^4 \tilde{B}_2 
+ \frac{2\alpha}{N} \nabla_k \left[ \nabla^k N 
\left( \nabla^2 \tilde{B}_2 - \tilde{B}_3 \right) \right]
- \left( \beta R + \alpha a_k a^k \right) \nabla^2 \tilde{B}_2
\nonumber
\\ &&
- 2 \alpha \left( \nabla^2 \tilde{B}_3 
- \frac{\nabla^2 N}{N} \tilde{B}_3 \right) \,,
\label{eqmultipliers}  
\end{eqnarray}
where $\tilde{B}_{2,3} \equiv B_{2,3}/N$. As we commented before, the equation of motion associated to $\delta N$ is homogeneous in the Lagrange multipliers $B_{2,3}$.

It is illustrative ot make a comparison with the field equations of general relativity in the ADM formalism. Indeed, it turns out that the effective theory shown above coincides with the dynamics of general relativity under a specific limit, and in particular for the asymptotically flat configurations. The limit consists in setting the coupling constants $\beta = 1$ and $\alpha = 0$, while $\omega$ is irrelevant in $M_{\mbox{\tiny AF}}$ as already discussed. In addition, the Lagrange multipliers $B_{1,2,3}$ are turned off, which automatically solves Eq.~(\ref{eqmultipliers}). Since we are dealing with the $M_{\mbox{\tiny AF}}$ subspace, we know that the constraints $\tilde{\mathcal{H}}, \tilde{\phi}^1,\tilde{\phi}^2$ constraints of the kcH-theory holds. Thus, the set of constraints takes the form in this limit
\begin{eqnarray}
&&
\mathcal{H}^i = - 2 \nabla_k \pi^{ik} = 0\,,
\label{grmomentum}
\\ &&
\mathcal{H} = 
\frac{1}{\sqrt{g}} \pi^{ij} \pi_{ij} - \sqrt{g} R 
= 0 \,,
\label{grhamiltonian}
\\ &&
\phi^1 = \pi = 0 \,,
\label{grpicero}
\\ && 
\phi^2 =
{\nabla^2 N} - NR  = 0 \,.
\label{grgaugefixing}
\end{eqnarray}
Equation (\ref{grmomentum}) is the momentum constraint of general relativity. Condition $\pi = 0$ (\ref{grpicero}) is admissible as a gauge fixing condition in the $M_{\mbox{\tiny AF}}$ subspace of general relativity. Equation (\ref{grgaugefixing}) is the condition neccesary for the time preservation of this gauge fixing condition \cite{DeWitt:1967yk}, and constraint (\ref{grhamiltonian}) is the usual Hamiltonian constraint of general relativity in the same gauge. The remainning field equations are the evolution equations (\ref{gpuntoz1free}) and (\ref{pipuntoz1free}), which take the form
\begin{eqnarray}
&&
\dot{g}_{ij} = 
\frac{2 N}{\sqrt{g}} \pi_{ij} + 2 \nabla_{(i} N_{j)}  \,,
\\ &&
\dot{\pi}^{ij} = 
- \frac{2 N}{\sqrt{g}}  
\left( \pi^{ik} \pi_k{}^j  
- \frac{1}{4} g^{ij}  \pi^{kl} \pi_{kl}  \right)
- 2 \nabla_k N^{(i} \pi^{j)k} + \nabla_k ( N^k \pi^{ij})
\nonumber
\\ &&
- \sqrt{g} N ( R^{ij} - \frac{1}{2} g^{ij} R )
+ \sqrt{g} ( \nabla^{ij} - g^{ij} \nabla^2 ) N \,.
\end{eqnarray}
These are the usual evolution equations of the ADM formulation of general relativity evaluated on the $\pi = 0$ gauge in $M_{\mbox{\tiny AF}}$. Therefore, in $M_{\mbox{\tiny AF}}$, in the limit $\beta = 1$ and $\alpha = 0$ and with the Lagrange multipliers $B_{1,2,3}$ turned off, the dynamics of this theory is exactly the same as the dynamics of general relativity when the gauge fixing condition $\pi = 0$ is imposed on the side of general relativity. This result is related to the connection found in Ref.~\cite{Bellorin:2010je} between the nonprojectable Ho\v{r}ava theory and general relativity (under an analogous limit for the coupling constants).

\paragraph{Gravitational waves.}
We study the linear-order perturbative version of the vacuum field equations of the previously shown large-distance effective theory. We define the Minkowski space, which is a solution of all the constraints and equations of motion, by $g_{ij} = \delta_{ij}$, $\pi^{ij} = 0$, $N = 1$ and the multipliers set to zero, $N_i = B_{1,23}= 0$. The perturbation of this solution is 
\begin{equation}
 g_{ij} = \delta_{ij} + h_{ij} \,,
 \quad \pi^{ij} = p_{ij} \,,
 \quad N = 1 + n \,,
 \quad N_i = n_i \,. 
\end{equation} 
As part of the ansatz we turn off the perturbation of the Lagrange multipliers $B_{1,2,3}$. At first sight it is not obvious that this choice is valid since we are dealing with the problem of extremizing the action, hence a consistent solution that involves all field variables must be found, as we commented previously (to be precise, what we pursue here is the wave equation for the independent propagating modes, which requires to solve consistently the rest of variables). At the end of the analysis we will see that the choice $B_{1,2,3}=0$ is consistent (part of the requisites is that the Eq.~(\ref{eqmultipliers}) is automatically solved).

We perform the usual transverse/longitudinal decomposition
\begin{equation}
h_{ij} =
h_{ij}^{TT} + \frac{1}{2} \left( \delta_{ij} - \frac{\partial_{ij}}{\Delta} \right) h^T + \partial_{(i} h_{j)}^L \,,
\label{decomposehij}
\end{equation}
where $\Delta \equiv \partial_{kk}$ is the flat Laplacian and these variables are subject to $h^{TT}_{kk} = \partial_k h^{TT}_{ki} = 0$. We apply the analogous decomposition on $p^{ij}$. We impose the transverse gauge $\partial_k h_{ki} = 0$, which fixes the gauge symmetry of spatial diffeomorphisms, hence in this gauge $h_i^L  = 0$.

We assume that all the perturbative variables that are fixed by elliptic equations, hence nonradiative, satisfy the asymptotically flat conditions (\ref{asympflat}). The linear-order momentum constraint $\mathcal{H}^i$ (\ref{momentumconstraintfin}) takes the form $\partial_i p^{ij} = 0$, hence the longitudinal sector of the canonical momentum vanishes, $p^L_i = 0$. Constraint $\phi^1$ (\ref{nablapi}) becomes $\Delta p^T = 0$. With the prescribed boundary conditions, the only solution of this equation is $p^T = 0$. The linear-order $\mathcal{H}$ and $\phi^2$ constraints become, respectively,
\begin{eqnarray}
&&
\Delta \left(\beta h^T +  2 \alpha n\right) = 0 \,,
\label{eqnh1}
\\ &&
\Delta^2 \left( \beta h^T + \gamma_2 n \right) = 0 \,. 
\label{cconstraintlinear}
\end{eqnarray}
Since $h^T$ and $n$ are of order $\mathcal{O}(r^{-1})$ asymptotically, the only solution to these equations is that the two combinations inside the brackets are zero. This is equivalent, whenever $\beta \neq 0$ and $\alpha \neq 2\beta$, to $h^T = n = 0$. Thus, the set of constraints, together with the transverse gauge, eliminate the longitudinal sector of $h_{ij}$ and $p_{ij}$, the scalars $h^T$ and $p^T$ and the lapse function $n$. So far the unfixed variables are the conjugate pair $(h_{ij}^{TT},p_{ij}^{TT})$ and the Lagrange multiplier $n_i$.

Next, we move to the evolution equations. Taking into account the already-fixed variables, the longitudinal sector of the linear-order Eq.~(\ref{gpuntoz1free}) yields the equation for $n_i$,
\begin{equation}
 \Delta n_i + \partial_i \partial_k n_k = 0 \,.
\end{equation} 
By taking the divergence of this equation, combined with the asympotic condition $n_i = \mathcal{O}(r^{-1})$, it is easy to see that this equation implies $n_i = 0$ everywhere. The transverse-traceless sectors of Eqs.~(\ref{gpuntoz1free}) and (\ref{pipuntoz1free}) yield
\begin{equation}
\dot{h}_{ij}^{TT} = 2p_{ij}^{TT} \,,
\quad
\dot{p}_{ij}^{TT} = \frac{\beta}{2} \Delta h_{ij}^{TT} \,.
\label{wavecan1}
\end{equation}
These equations imply the wave equation with wave speed $\sqrt{\beta}$,
\begin{equation}
\ddot{h}_{ij}^{TT} - \beta \Delta h_{ij}^{TT} = 0 \,.
\end{equation}
At this point we can see that having set the Lagrange multipliers $B_{1,2,3} = 0$ has leaded to consistent evolution equations for the independent propagating modes, whereas the rest of nonpropagating variables has been fixed consistently. We comment again about the time dependence of the lapse function: here we have obtained $n = 0$, this is a consequence of the linear order of the analysis. $n$ and $h^T$ acquire nonzero expressions for higher orders in perturbations (or in situations with sources), as happens in general relativity for the case of $h^T$ \cite{Arnowitt:1962hi}. Since $h^{TT}_{ij}$ and $p^{TT}_{ij}$ are in general waves that depend on time, $n$ and $h^T$ acquire an induced dependence on time at higher order in perturbations.

Besides the physical content, this perturbative analysis is a way to check explicitly that the $z=1$ theory is mathematically consistent, in the sense that all the constraints have been solved by the appropiated variables and the remaining canonical modes propagate with consistent wave equations. This confirms that there are two propagating physical modes.

\paragraph{Homogeneity and isotropy.}
In this section we take the large-distance effective theory for a cosmological application. We fix the symmetry of spatial diffeomorphisms by imposing the gauge $N_i = 0$. We consider the gravitational field coupled to a perfect fluid of density $\rho$ and pressure $P$. We adopt the point of view of Ref.~\cite{Carroll:2004ai}, where the nonrelativistic Eisntein-aether theory is coupled to a relativistic perfect fluid. It has the energy-momentum tensor,
\begin{equation}
T_{\mu\nu} =
8 \pi G \left( \rho u_\mu u_\nu
+ P (g_{\mu\nu} + u_\mu u_\nu) \right) \,,
\label{perfectfluid}
\end{equation}
where we have introduced the coupling constant $G$ as the gravitational constant weighting the coupling of the gravitational theory to sources. We set the four-velocity of the fluid as $u^\mu = ( N^{-1} , 0 , 0 , 0 )$, which correspond to a fluid at rest. Since the perfect fluid is modeled in terms of $T_{\mu\nu}$ rather than a Lagrangian, we need a criterium to couple $T_{\mu\nu}$ to the field equations. This can be achieved by demanding that the variations with respect to the canonical variables acquire contributions given by the components of $T_{\mu\nu}$. To follow a standard reference, we write the canonical (ADM) field equations of general relativity coupled to the perfect fluid under the same physical considerations (without the $\pi = 0$ gauge),
\begin{eqnarray}
&&
\nabla_k \pi^{ik} = 0\,,
\label{grmomentumsourced}
\\ &&
\frac{1}{\sqrt{g}} 
 \left( \pi^{ij} \pi_{ij} - \frac{1}{2} \pi^2 \right) 
 - \sqrt{g} R 
= - 16 \pi G \sqrt{g} \rho \,,
\label{grhamiltoniansourced}
\\ &&
\dot{g}_{ij} = 
\frac{2 N}{\sqrt{g}} 
 \left( \pi_{ij} - \frac{1}{2} g_{ij} \pi \right) +  
2 \nabla_{(i} N_{j)}  \,,
\\ &&
\dot{\pi}^{ij} = 
- \frac{2 N}{\sqrt{g}}  
\left[ \pi^{ik} \pi_k{}^j - \frac{1}{2} \pi \pi^{ij}   
- \frac{1}{4} g^{ij} 
  \left( \pi^{kl} \pi_{kl} - \frac{1}{2} \pi^2 \right) \right] 
- 2 \nabla_k N^{(i} \pi^{j)k} + \nabla_k ( N^k \pi^{ij})
\nonumber
\\ &&
- \sqrt{g} N ( R^{ij} - \frac{1}{2} g^{ij} R )
+ \sqrt{g} ( \nabla^{ij} - g^{ij} \nabla^2 ) N 
+ 8\pi G \sqrt{g} g^{ij} N P \,.
\label{pipuntogrsource}
\end{eqnarray}
The Hamiltonian constraint (\ref{grhamiltoniansourced}) and the $\dot{\pi}^{ij}$ equation (\ref{pipuntogrsource}) get contributions from the source since they are obtained by varying $N$ and $g_{ij}$ respectively. Thus, in the theory we study here we establish the following rules: the variation with respect to $N$ adds $-16 \pi G \sqrt{g} \rho$ and the variation with respect to $g_{ij}$ adds $8\pi G \sqrt{g} N g^{ij} P$, whereas the variation with respect to $\pi^{ij}$ does not get contribution from the source. In this way, constraints $\mathcal{H}$ and $\phi^2$ of the theory get contributions from the source,
\begin{eqnarray}
  &&
 \mathcal{H} = 
 \frac{1}{\sqrt{g}} \left( \pi^{ij} \pi_{ij} - \omega \pi^2 \right) 
 - \sqrt{g} \left( \beta R + \alpha a_k a^k  - 2\alpha \frac{\nabla^2 N}{N}  \right) + 16 \pi G \sqrt{g} \rho \,,
 \label{hamiltonianconstsourced}
 \\ &&
 \phi^2 = 
 N^{-1} \nabla^2 \left[ \mathcal{E} 
 - 12 \pi G \sqrt{g} N ( \rho - P ) \right]
 \,,
 \label{c2z1}
\end{eqnarray}
where $\mathcal{E}$ is given in (\ref{phi2z1}), whereas constraints $\mathcal{H}^i$ and $\phi^1$ maintain their forms given in Eqs.~(\ref{momentumconst}) and (\ref{phiuno}). With regard to the evolution equations, we may simplify the exposition thanks to the fact that we are going to consider the cosmological configuration only with vanisihing Lagrange multipliers $B_{1,2,3}$. As we commented above, due to the Hamiltonian constraint $\mathcal{H}$, the equation derived by taking variations of the action with respect to $N$ is always homogeneous in $B_{1,2,3}$, even in the presence of sources. Hence this equation is automatically solved by $B_{1,2,3} = 0$, such that we do not longer consider it. The equation derived from $\delta g_{ij}$, with all the Lagrange multipliers turned off, takes the form
\begin{eqnarray}
 &&
 \dot{\pi}^{ij} = 
 - \frac{2 N}{\sqrt{g}}  
 \left[ \pi^{ik} \pi_k{}^j - \omega \pi \pi^{ij} 
 - \frac{1}{4} g^{ij} \left( \pi^{kl} \pi_{kl} - \omega \pi^2 \right) \right]
 \nonumber
 \\ &&
 - \sqrt{g} \left[
 \beta ( R^{ij} - \frac{1}{2} g^{ij} R )
 + \alpha ( a^i a^j - \frac{1}{2} g^{ij} a_k a^k ) 
 - \beta ( \nabla^{ij} - g^{ij} \nabla^2 ) \right] N
 \nonumber
 \\&& 
 + 8\pi G \sqrt{g} g^{ij} N P \,.
 \label{pipuntosourced}
\end{eqnarray}
The equation derived from the $\delta \pi^{ij}$ variation does not get contribution from the source, it is the same Eq.~(\ref{gpuntoz1free}).

Now we consider a flat homogeneous and isotropic configuration. The spatial metric can be casted in the form $g_{ij} = a(t)^2 \delta_{ij}$. The fields $N$ and $\pi^{ij}$, as well as the sources $\rho$ and $P$, are regarded as functions only of time. Since $N$ is a function only of time, we may use the symmetry of reparameterizing the time to set $N = 1$ (other choices would leave active the time dependence of $N$). The last part of our ansatz is that, for concretness, we look for solutions with $B_{1,2,3} = 0$, as we anticipate. With these settings the constraints $\mathcal{H}^i$ (\ref{momentumconst}), $\phi^1$ (\ref{phiuno}) and $\phi^2$ (\ref{c2z1}) are automatically solved since in all of them there are spatial derivatives acting on pure functions of time. The equation of motion (\ref{gpuntoz1free}) can be solved completely for $\pi^{ij}$, yielding
\begin{equation}
 \pi^{ij} = 
 \left(\frac{1}{1 - 3\omega}\right) \dot{a} \delta_{ij} \,.
\end{equation}
By inserting all this information in the $\mathcal{H}$ constraint (\ref{hamiltonianconstsourced}) and the equation of motion (\ref{pipuntosourced}), we obtain that they become a system of equations of the kind of the Friedmann equations, namely
\begin{equation}
 \left( \frac{\dot{a}}{a} \right)^2 = 
 \frac{8\pi G_c}{3} \rho \,,
 \quad
 \frac{\ddot{a}}{a} =
 - \frac{4\pi G_c}{3}  ( \rho + 3 P ) \,,
\end{equation}
where $G_c \equiv 2 (3\omega - 1) G$ plays the role of effective gravitational constant of cosmological scale. 

This analysis may be continued by incorporating cosmological bounds on the relevant coupling constants, a study that we leave for future work. We give in advance that the constant $\omega$ is still free, and in the analysis on some observational bounds that we will do in the next section it will remain unaffected. Hence we expect a good adaptability of this theory to the cosmological observations. In addition we recall that here we have not considered the cosmological constant and we have restricted the homogeneous and isotropic ansatz to flat geometries.

\paragraph{Observational bounds.}
We perform a quick analysis directed to some of the gravitational phenomena whose results can be applied directly to this theory at the stage we have analyzed it, without requiring further major analysis. These are the speed of gravitational waves and the PPN parameters. The former is direct since we have deduced the propagating equations at the leading order and the latter is direct thanks to the correspondence this theory has with the kcH-theory. We consider the observational bounds only in the $z=1$ truncation, since it is the effective theory for large distances.

We recall that on the theoretical side the available coupling constants are $G$, $\omega$, $\beta$ and $\alpha$. Previously we showed that the theory propagates the same tranverse-traceless modes that linearized General Relativity does (in the transverse gauge), but with speed $c_T = \sqrt{\beta}$. The recent obervation \cite{TheLIGOScientific:2017qsa} of a gravitational wave associated with a electromagnetic signal put the speed of the gravitational waves extremely close to the speed of light. The bounds are \cite{Monitor:2017mdv}
\begin{equation}
- 3 \times 10^{-15} \leq c_T - 1 \leq 7 \times 10^{-16} \,.
\end{equation}
In this theory this extremely narrow window is safely satisfied by putting $\beta = 1$.

The second analysis is given by the PPN parameters valid for solar-system phenomena. This kind of observational bound applies for asymptotically flat configurations. We have seen that the asymptotically flat configurations of this theory are exactly the same of the kcH-theory, hence we may take the results of the kcH-theory. For the nonprojectable $z=1$ Ho\v{r}ava theory the PPN coefficients were obtained in Ref.~\cite{Blas:2011zd}. Those authors used the generally-covariant version of the theory that introduces a gauge scalar field. The resulting covariant theory coincides with the hypersurface-orthogonal Einstein-aether theory, and it is also called the khronometric theory \cite{Jacobson:2000xp,Blas:2009ck,Jacobson:2010mx,Foster:2005dk}. In turn, the PPN coefficients of the generic nonprojectable Ho\v{r}ava theory found in \cite{Blas:2011zd} can be adapted for the kcH-theory \cite{Bellorin:2016hcu}. The result for the kcH-theory, and hence for this theory, is that the PPN parameters coincide with the values of General Relativity, except for
\begin{equation}
\alpha_2^{\mbox{\tiny PPN}} =
\frac{1}{8} \alpha_1^{\mbox{\tiny PPN}} =
\beta - 1 - \frac{\alpha}{2}  \,. 
\end{equation}
The current stringtest bound is $\alpha_2^{\mbox{\tiny PPN}} < 10^{-9}$ \cite{Will:2014kxa}. Since $\beta$ has already been set equal to $1$, it seems that in order to satisfy this bound safely the better choice is to set $\alpha = 0$. We remark that this and the $\beta = 1$ condition can be implemented without modifying the essential physical features of this theory. In particular, condition $\alpha = 0$ drops the $a_i a^i$ term out from the $z=1$ potential. In this theory this terms is not crucial for the stability of any extra mode, because there is not extra mode. Indeed, in the perturbative analysis we solved the constraints explicitly and got the propagating field equations. The condition $\alpha = 0$ is not in contradiction with the procedure used there. Moreover, in the phenomenological criteria the coupling constants of the terms of higher order derivatives are left completely unaffected.




\end{document}